\documentclass{article}
\usepackage{graphicx,epsfig}
\newcommand{\ccc}[1]{} 

\begin{document}
\title{Theoretical aspects of high energy elastic nucleon scattering}

\author{{\slshape {Vojt\v{e}ch Kundr\'{a}t$^1$, Jan Ka\v{s}par$^{2,1}$},
Milo\v{s} Lokaj\'{i}\v{c}ek$^1$}\\[1ex]
$^1$Institute of Physics of the AS CR, v.v.i., 182 21 Prague 8, \\
Czech Republic\\
$^2$CERN, 1211 Geneve 23, Switzerland}



\maketitle

\begin{abstract}
The eikonal model must be denoted as strongly 
preferable for the analysis of elastic high-energy 
hadron collisions. The given approach allows to derive 
corresponding impact parameter profiles that characterize 
important physical features of nucleon collisions, e.g., 
the range of different forces. The contemporary phenomenological 
analysis of experimental data is, however, not able to determine 
these profiles unambiguously, i.e., it cannot give the answer 
whether the elastic hadron collisions are more central or more 
peripheral than the inelastic ones. However, in the collisions 
of mass objects (like protons) the peripheral behavior of 
elastic collisions should be preferred.
\end{abstract}
\section{Coulomb-hadronic interference}
\label{sec1}
The first attempt to determine the complete elastic scattering
amplitude $F^{C+N}(s,t)$ for high energy collisions of 
charged nucleons has been made by Bethe \cite{beth}:
\begin{equation}
F^{C+N}(s,t) = F^C(s,t) e^{i \alpha \Psi(s,t)} + F^N(s,t)
\label{wy1}
\end{equation}
where $F^C(s,t)$ is Coulomb amplitude (known 
from QED), $F^N(s,t)$ - elastic hadronic amplitude,
$\alpha \Psi(s,t)$ - real relative phase between Coulomb 
and hadronic scattering and $\alpha=1/137.036$ is the 
fine structure constant. This relative phase has been 
specified by West and Yennie \cite{west} using the 
Feynman diagram technique (one photon exchange) as 
\begin{equation}
\Psi_{WY}(s,t)=-\ln{-s\over t}-\int_{-4p^2}^{0}{dt'\over |t'-t|}
\left[1-{F^N(s,t')\over F^N(s,t)}\right];
\label{wy2}
\end{equation}
$p$ representing the value of the incident momentum
in the centre-of-momentum system. By their construction
(see Eq. (\ref{wy2}), the phase $\Psi_{WY}(s,t)$ is to 
be real. 

However, it has been shown in Ref. \cite{kun1}
that this requirement can be fulfilled only provided 
the phase of the elastic hadronic amplitude $\zeta^N(s,t)$
defined in our case as
\begin{equation}
F^{N}(s,t) = i |F^{N}(s,t)| e^ {-i \zeta^{N} (s,t)},
\label{nu1}
\end{equation}
is $t$ independent at all kinematically allowed 
values of $t$. Rigorous proof has been given for
$|\zeta^N(s,t)| < \pi$, which is fulfilled practically 
in all standard phenomenological models leading to the
central distribution of elastic processes in impact
parameter space. It is not yet clear if it holds also 
for $|\zeta^N(s,t)| < 2 \pi$ corresponding to peripheral 
behavior; see Fig. 1 where both the types of phase $t$ 
dependences are represented. The corresponding $t$ 
dependences of imaginary parts of complex relative phases 
$\alpha \Psi_{WY}(s,t)$ are shown in Fig. 2.
\vspace*{1cm}
\begin{figure}[h!]
\centerline{
\begin{tabular}{cc}
\includegraphics[width=0.5\textwidth]{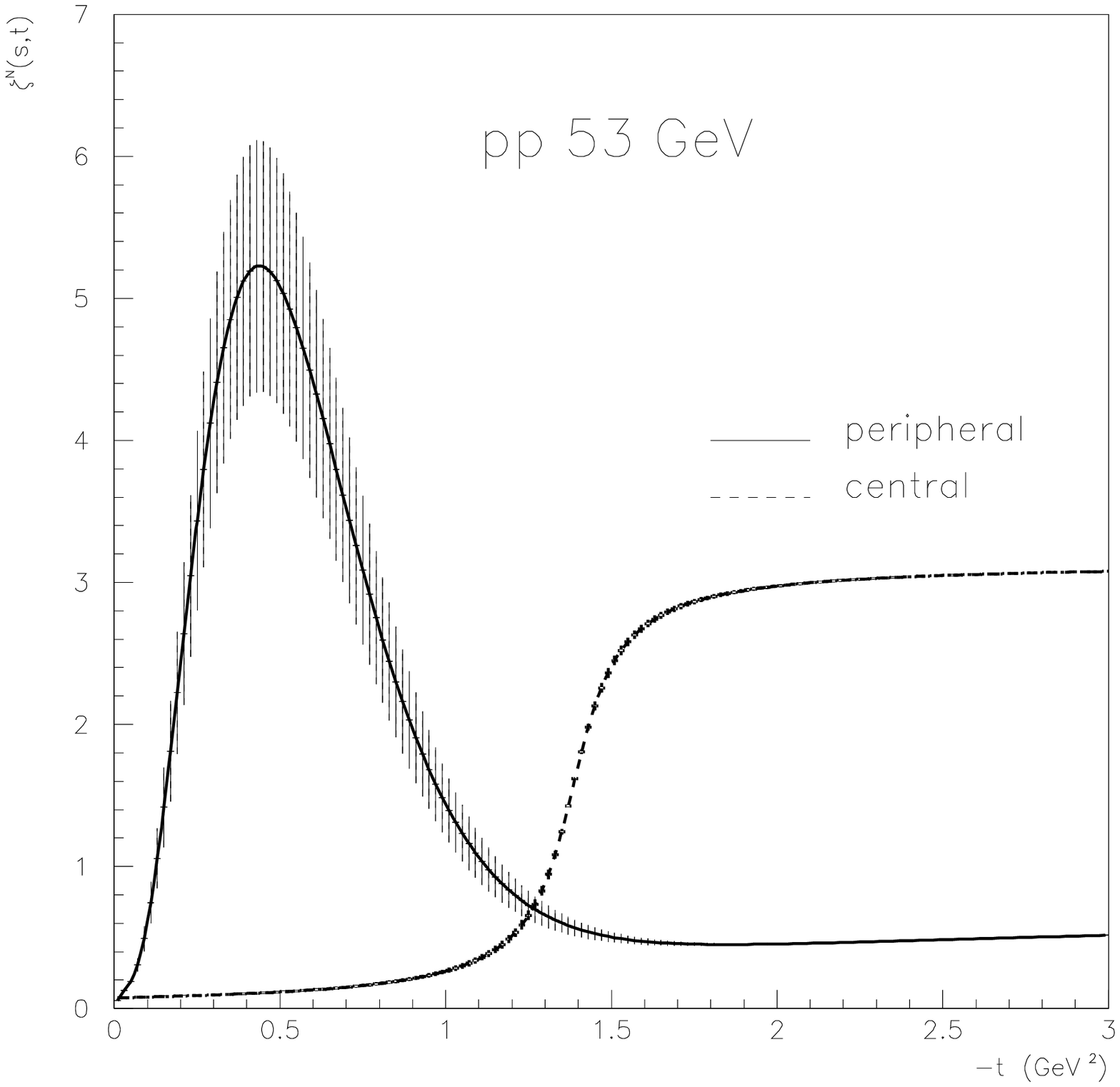} & 
\includegraphics[width=0.5\textwidth]{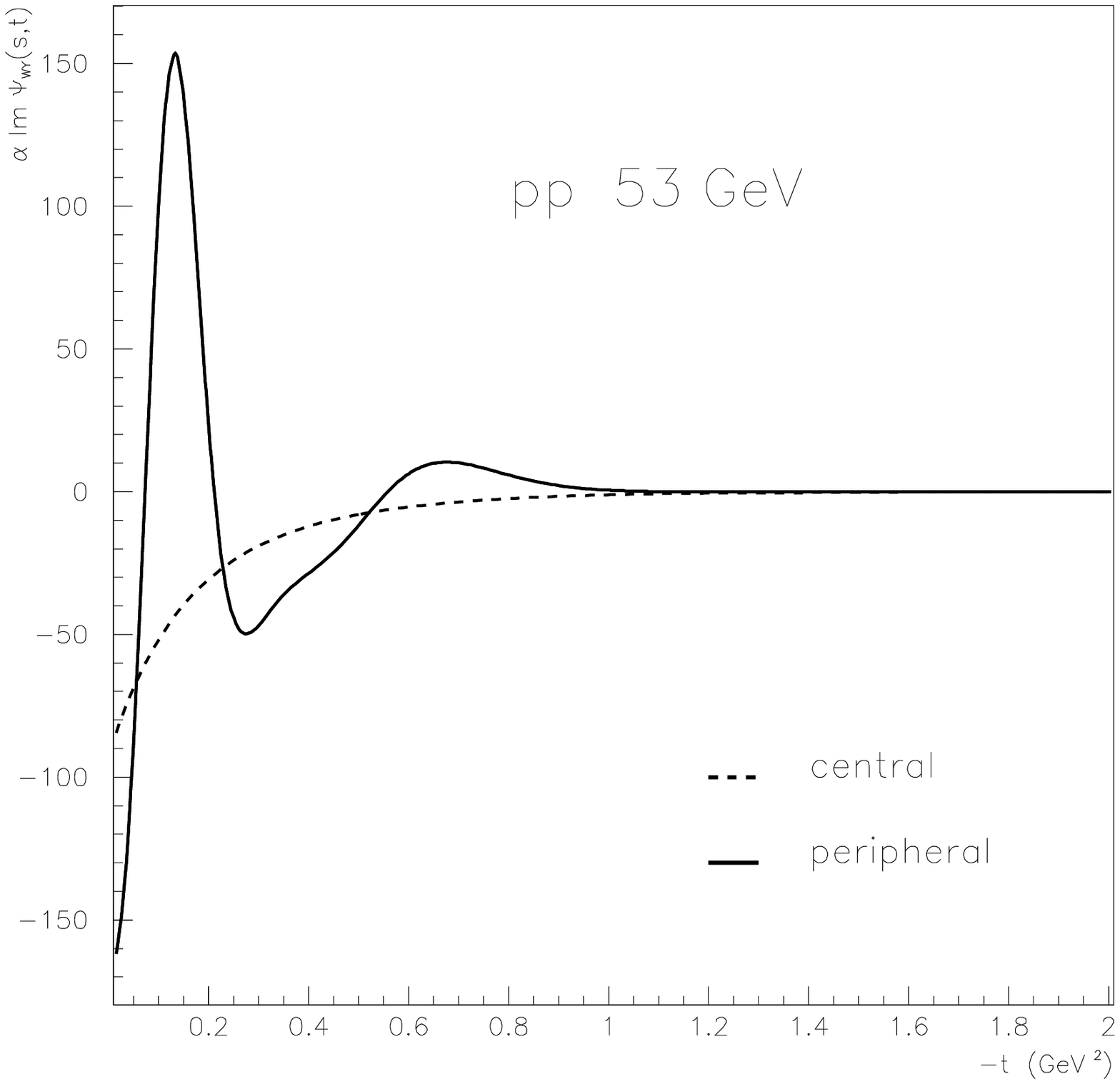}\\[-0.7cm]
\begin{minipage}[t]{0.45\textwidth}
\vspace*{-1.5cm}
\caption{\label{fig:kundrat_vojtech_eds09_fig1}
$t$ dependences of the hadronic phases $\zeta^N(s,t)$ leading
to the central and peripheral behaviors of elastic $pp$ scattering.} 
\end{minipage} & \begin{minipage}[t]{0.45\textwidth} 
\vspace*{-1.5cm}
\caption{\label{kundrat_vojtech_eds09_fig2}$t$ dependences of imaginary 
parts of the complex WY phases $\alpha \Psi_{WY}(s,t)$ corresponding to 
hadronic phases from Fig. 1.} 
\end{minipage} \\
&\\[-1.0cm]
\end{tabular}}
\end{figure}
\vspace*{0.5cm}

There is, however, no reason why the elastic hadronic phase 
should be independent of $t$ variable at all measured values 
of $t$. Thus the other approach different from the West and 
Yennie formalism and based on the eikonal approach should be 
preferred - see, e.g.,  Ref. \cite{kun2}. In such a case 
the complete elastic scattering amplitude $F^{C+N}(s,t)$
is related to the complete eikonal $\delta^{C+N}(s,b)$
with the help of Fourier-Bessel (FB) transformation
\begin{equation}
F^{C+N}(s,q^2=-t)= {s\over {4 \pi i}} \int\limits_{\Omega_b}d^2b
e^{i\vec{q} . \vec{b}} \left[e^{2i\delta^{C+N}(s,b)}-1\right],
\label{eik1} 
\end{equation}
where $\Omega_b$ is the two-dimensional Euclidean space of 
the impact parameter $\vec b$. Mathematically consistent 
use of FB transformation requires, of course, the existence 
of reverse transformation. However, at finite energies the 
amplitude $F^{C+N}(s,t)$ is defined in a finite region of 
$t$ only. Thus the consistent application of formula 
(\ref{eik1}) requires to take into account also the values of 
elastic amplitude from unphysical region where the elastic 
hadronic amplitude is not defined; for details see Refs.
\cite{adac}. This issue has been discussed by Islam 
\cite{isla} who has shown that the problem may 
be solved in a unique way by continuing analytically 
the elastic hadronic amplitude $F^N(s,t)$ from 
physical to unphysical region of $t$ ; see also 
Ref. \cite{kun3}.

The common influence of both the Coulomb and elastic 
hadronic scattering then can be described by complete 
eikonal which is formed by the sum of both
the Coulomb $\delta^C(s,b)$ and hadronic $\delta^N(s,b)$ 
eikonals. Then the complete amplitude (valid at any $s$ 
and $t$) can be finally written \cite{kun2} as
\begin{equation}
F^{C+N}(s,t) = \pm {\alpha s\over t}f_1(t)f_2(t) +
F^{N}(s,t)\left [1\mp i\alpha G(s,t) \right ],  
\label{kl1}
\end{equation}
where
\begin{equation}
G(s,t) = \int\limits_{-4p^2}^0
dt'\left\{ \ln \left( {t'\over t} \right )
{d \over{dt'}}
\left[f_1(t')f_2(t')\right]
+ {1\over {2\pi}}\left [{F^{N}(s,t')\over F^{N}(s,t)}-1\right]
I(t,t')\right\},
\label{kl2}
\end{equation}
and
\begin{equation}
I(t,t')=\int\limits_0^{2\pi}d{\Phi^{\prime \prime}}
{f_1(t^{\prime \prime})f_2(t^{\prime \prime})\over t^{\prime \prime}}, \;\;
t^{\prime \prime}=t+t'+2\sqrt{tt'}\cos{\Phi}^{\prime \prime}.
\label{kl3}
\end{equation}
Here the two form factors $f_1(t)$ and $f_2(t)$ reflect the spatial
structure of colliding nucleons and should describe it in
a sufficiently broad interval of $t$.
As the Coulomb amplitude $F^C(s,t)$ is known from QED
the complete amplitude is determined practically by
the $t$ dependence of the hadronic amplitude $F^N(s,t)$.

\section{Impact parameter picture of elastic nucleon scattering}
\label{sec2}

As it has been already mentioned  
the mathematically consistent use of FB 
transformation introducing the impact parameter 
representation $h_{el}(s,b)$ of elastic scattering 
amplitude $F^N(s,t)$ requires its definition also 
in the unphysical region of $t$, i.e., for 
$t < t_{min}= -s+4m^2$ ($m$ being the nucleon 
mass). The function $h_{el}(s,b)$ must be, 
therefore, subdivided into two parts 
\cite{adac,isla}
\begin{eqnarray}
\hspace*{-0.8cm}h_{el}(s,b) &=& h_1 (s,b) + h_2 (s,b) =  \\
\label{hn2}
&=&{1 \over {4 p \sqrt{s}}} 
\int \limits_{t_{min}}^{0} \!\! dt \; F^{N}(s,t)\; J_{0}(b \sqrt{-t})
\label{el1} + {1 \over {4 p \sqrt{s}}} \! 
\int \limits_{-\infty}^{t_{min}} \! dt \; 
F^{N}(s,t) \; J_{0}(b \sqrt{-t}).
\nonumber 
\end{eqnarray}
Similar expressions can be obtained also for the impact 
parameter representation $g_{inel}(s,b)$ of the
inelastic overlap function $G_{inel}(s,t)$ introduced
in Ref. \cite{hove}.

The unitarity equation in the impact parameter space
can be then written as \cite{adac,isla}
\begin{equation}
\Im h_{1}(s, b) = |h_{1}(s,b)|^2 + g_{1}(s, b) + K(s, b)
\label{ue1}
\end{equation}
where the correlation function $K(s,b)$ is very small
compared to the other functions appearing in 
Eq. (\ref{ue1}) \cite{kun3}. 

The total cross section and integrated elastic and inelastic
cross sections then equal to
\begin{equation}
    \sigma_{tot}(s) = 8 \pi  \!\!
\int \limits_{0}^{\infty} \!\! b d b \; \Im h_{1}(s, b);
\hspace*{0.1cm}    \sigma_{el}(s) = 8 \pi  \!\!
\int \limits_{0}^{\infty} b d b \;  |h_{1}(s, b)|^2;
\hspace*{0.1cm}
     \sigma_{inel}(s) = 8 \pi \!\!
\int \limits_{0}^{\infty} b d b \; g_{1}(s, b).             
\label{cs1} 
\end{equation}
Eqs. (\ref{cs1}) are valid provided 
\begin{equation}  
\int \limits_{0}^{\infty} b d b \; \Im h_{2}(s, b) \; = \; 0,
\hspace*{1cm}
\int \limits_{0}^{\infty} b d b \; g_{2}(s, b)\; = \; 0.   
\label{ps1} 
\end{equation}

The functions $\Im h_1(s,b)$, $g_1(s,b)$ and $|h_1(s,b)|^2$  
represent then the impact parameter profiles; they describe
the density of interactions between two colliding nucleons
in dependence on their mutual distance. The first two oscillate
at grater values of $b$, but their mean squared values
characterizing the mean ranges of corresponding forces 
acting between the colliding particles can be established 
directly from the elastic hadronic amplitude $F^N(s,t)$ 
\cite{kun4}. For the mean squared value of elastic impact 
parameter it has been derived 
\begin{equation}
{\langle } b^2(s){\rangle }_{el}  \; =  \;
 4 \; {{\int \limits_{t_{min}}^{0}\!\! dt \;|t| \;
\Bigl({d \over{dt}} |F^{N}(s,t)|\Bigr)^2}
\over { \int \limits_{t_{min}}^{0}\!\! dt \; |F^{N}(s,t)|^2}} +
4 \; {{\int \limits_{t_{min}}^{0}\!\! dt \;|t| \;
|F^{N}(s,t)|^2 \Bigl( {d \over {dt}} \zeta^{N}(s,t)\Bigr)^2} 
\over { \int \limits_{t_{min}}^{0} \!\!dt \; |F^{N}(s,t)|^2}}.
\label{rmse}
\end{equation}
Similarly the total and inelastic mean squared values equal to
\begin{equation}
{\langle }b^2(s){\rangle }_{tot}\;\; = 2 B(s,0); \hspace*{0.3cm}
{\langle } b^2(s){\rangle }_{inel}\;\; = \; 
{{\sigma_{tot}(s)} \over {\sigma_{inel}(s)}} {\langle }  
b^2(s){\rangle }_{tot} \; - \;
{{\sigma_{el}(s)} \over {\sigma_{inel}(s)}} {\langle } b^2(s){\rangle
}_{el}.
\label{rmst}
\end{equation}
Here the diffractive slope is defined as
\begin{equation}
B(s,t)= {d\over {dt}}\left[\ln {d \sigma^{N}\over {dt}}\right] =
{2\over |F^{N}(s,t)|}{d\over {dt}}|F^{N}(s,t)|.
\label{sl1}
\end{equation}

\section{Impact parameter profiles for $pp$ scattering at 53 GeV}
\label{sec3}
Basic results concerning the analysis of $pp$ 
elastic scattering data at energy of 53 GeV at
the ISR \cite{land} based on the eikonal approach
have been published in Ref. \cite{kun2}. Here we 
will mention only the results related 
to the impact parameter profiles.

In the quoted paper we have used the formulas
(\ref{kl1})-(\ref{kl3}) for the complete elastic
scattering amplitude $F^{C+N}(s,t)$ generating the
differential cross section 
\begin{equation}
{{ d \sigma_{el} } \over {dt}} =  {{\pi} \over {s p^2}} 
|F^{C+N}(s,t)|^2.
\label{ds1}
\end{equation}
The elastic hadronic amplitude, i.e., its modulus and
phase defined in Eq. (\ref{nu1}), has been conveniently 
parameterized in order to describe the $pp$ elastic scattering 
as central as well as peripheral process. While the $t$ 
dependence of the modulus can be almost unambiguously
determined from the data the phase can be only partially 
constrained. Both the possible alternatives (central and 
peripheral) have been presented in Ref. \cite{kun2}. 
The $t$ dependences of the corresponding shapes of the 
hadronic phase $\zeta^N(s,t)$ are shown in Fig. 1. 

Once the elastic hadronic amplitude $F^N(s,t)$ has been 
specified it has been possible to determine corresponding
impact parameter profiles together with their statistical 
errors with the help of FB transformation. And it has been 
also possible to determine the RMS values of the total, elastic 
and inelastic profiles with the help of Eqs. (\ref{rmse})
and (\ref{rmst}) for the central as well as the peripheral 
pictures of elastic $pp$ scattering. Their shapes corresponding 
to peripheral behavior are shown in Fig. 3 (for the central
case see Ref. \cite{kun1}); all RMS values are included 
in Table 1. In the central picture the elastic RMS is much 
lower than the inelastic one. This result agrees with the 
result of Miettinen \cite{miet}. It means that the protons 
in 'head-on' collisions 
\vspace*{1cm}
\begin{figure}[h!]
\centerline{
\begin{tabular}{cc}
\includegraphics[width=0.5\textwidth]{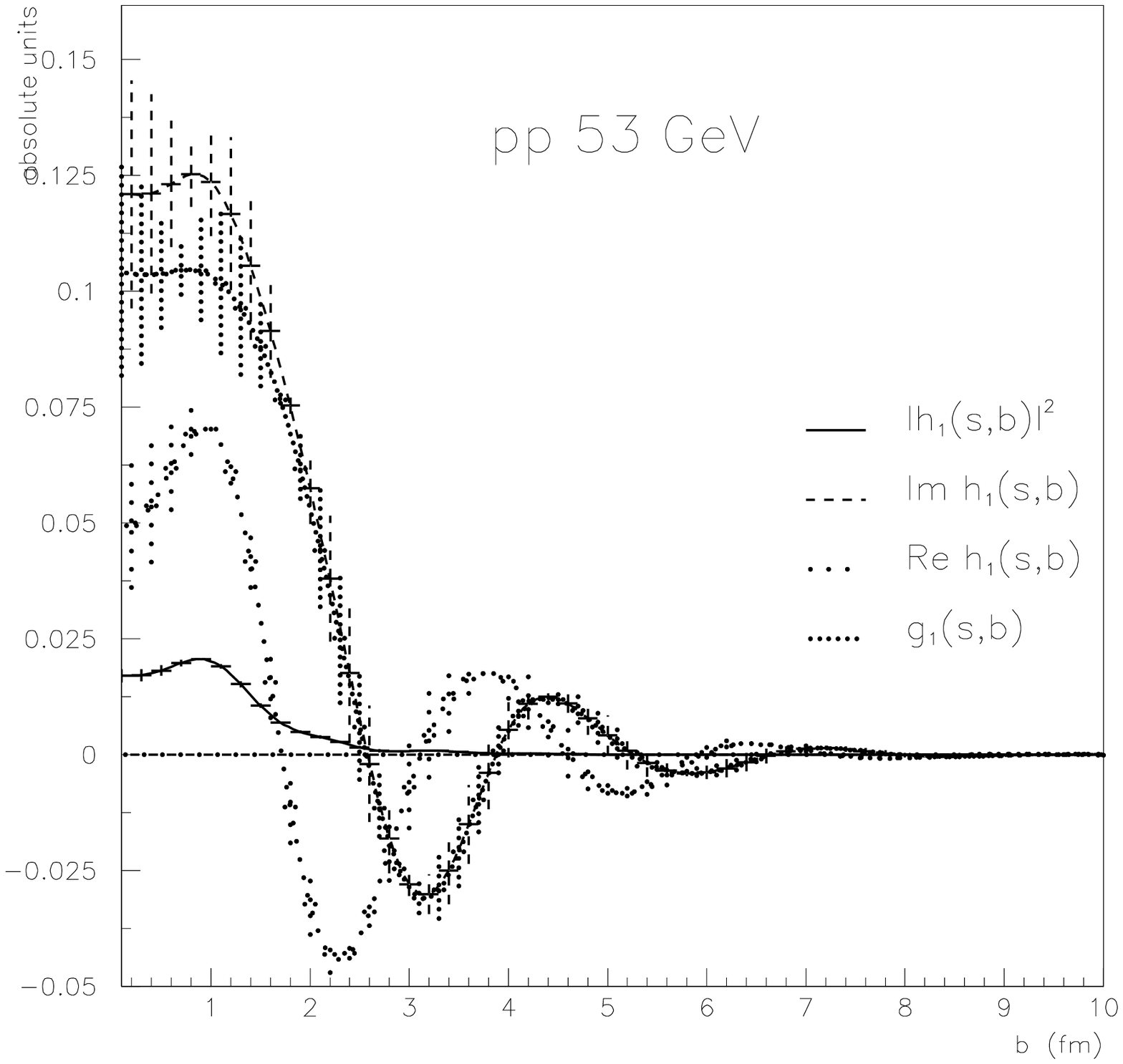} & 
\includegraphics[width=0.5\textwidth]{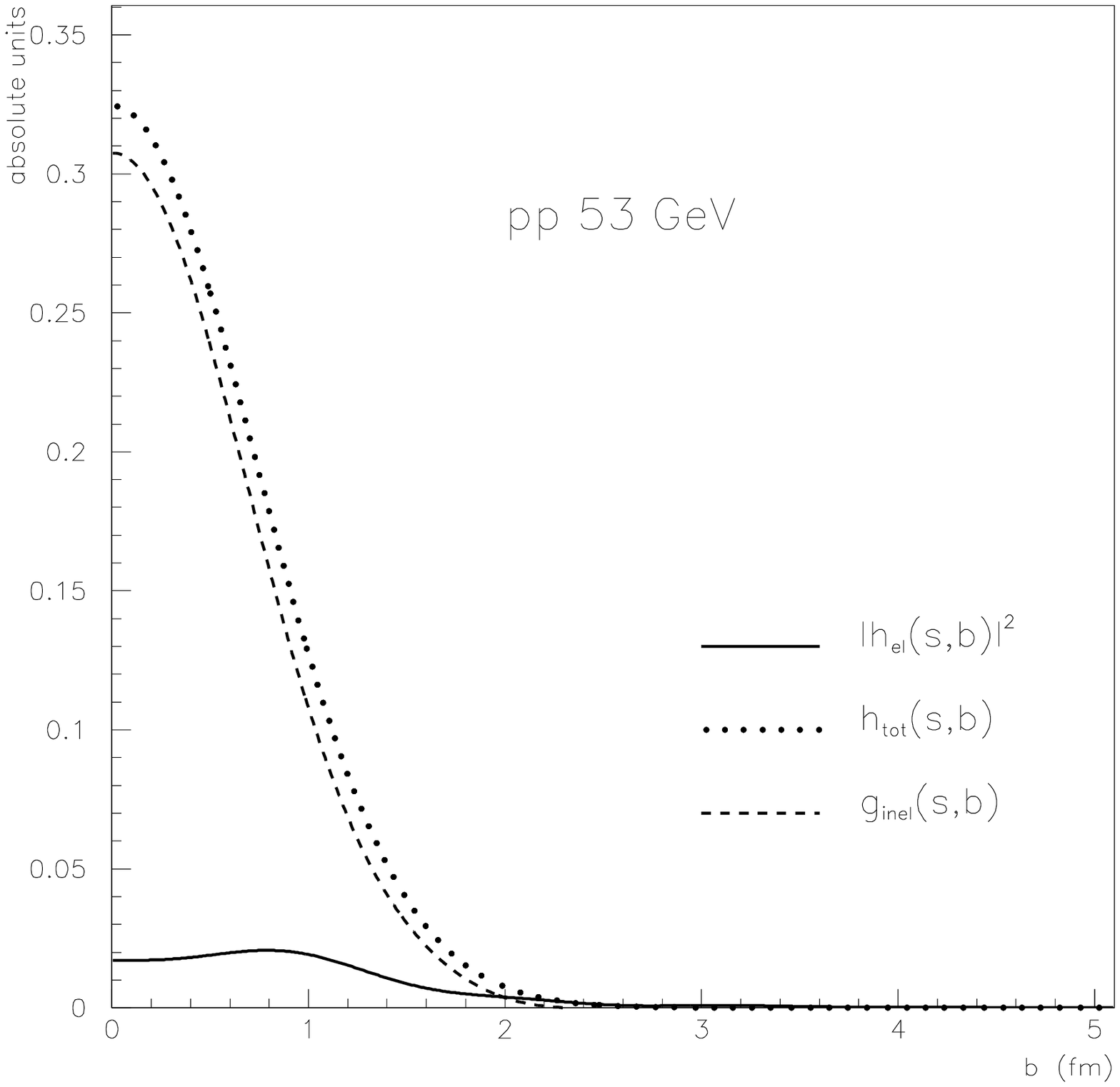}\\[-0.7cm]
\begin{minipage}[t]{0.45\textwidth}
\vspace*{-1.5cm}
\caption{\label{fig:kundrat_vojtech_eds09_fig3}
Oscillating peripheral profiles of elastic $pp$ scattering at 53 GeV.} 
\end{minipage} & \begin{minipage}[t]{0.45\textwidth} 
\vspace*{-1.5cm}
\caption{\label{kundrat_vojtech_eds09_fig4}
Positive peripheral profiles of elastic $pp$ scattering at 53 GeV.}  
\end{minipage} \\
&\\[-1.0cm]
\end{tabular}}
\end{figure}
should be rather transparent. 
In the peripheral case it is the second term in Eq. 
(\ref{rmse}) that gives the significant contribution 
to the elastic RMS value. Some profiles corresponding 
to the peripheral case exhibit greater oscillations 
at higher impact parameter values $b$. However, the 
oscillations can be removed as it will be shown 
in Ref. \cite{kun3}. It will be briefly described 
in the following.

It is necessary to construct actual (non-negative)
$pp$ profiles which correspond to the values of RMS 
derived with the help of Eqs. (\ref{rmse}) and 
(\ref{rmst}). The way how to do it is described  
in Refs. \cite{kun3,kun4}. It consists in adding an 
real function $c(s,b)$ to the both sides 
of the unitarity equation (\ref{ue1}).  
All dynamical characteristics corresponding 
to elastic hadron scattering will be preserved
if the function $c(s,b)$ appearing in Eqs. (\ref{pr1})
fulfills some additional conditions.

Unitarity condition (\ref{ue1}) may be then expressed as
\begin{equation}
\hspace*{-0.8cm}
h_{tot}(s,b)= |h_1(s,b)|^2 + g_{inel}(s,b)       
\label{pr2}
\end{equation}
where 
\begin{equation}
\hspace*{-0.8cm}
h_{tot}(s,b)=\Im h_1(s,b) + c(s,b), \;\;\;\;
g_{inel}(s,b) = g_1(s,b) + K(s,b) + c(s,b) 
\label{pr1}
\end{equation}
are non-negative functions for all $b$. The function 
$h_{tot}(s,b)$ must be positively semidefinite (and 
monotony decreasing function) at all values of $b$. 
The oscillating function $c(s,b)$ is required to 
cancel the oscillations from the total and inelastic 
profiles. The elastic peripheral profile $|h_1(s,b)|^2$
will remain unchanged. 
And integrated inelastic cross section is preserved 
if the function $c(s,b)$ fulfills the conditions
\begin{equation}
\int \limits_{0}^{\infty} b\; db \; c(s,b) = 0; \;\;\;\;\;\;\;\;\;
\int \limits_{0}^{\infty} b^3\; db\; c(s,b) \; = \; 0.
\label{rs1}
\end{equation}
These two conditions represent the integral conditions 
limiting the shape of the function $c(s,b)$. According
to Islam's approach \cite{isla} this function may be 
identified with the $\Im h_2(s,b)$ for which the conditions
(\ref{rs1}) are to be required. However, in the standard approach
the function $c(s,b)$ can be hardly determined analytically.
The best way at the present seems to specify it in a numerical 
way as it will be illustrated in the following. 
\begin{table}
\begin{center}
\caption{Root-mean-squares of impact parameters  
for $pp$ collisions at 53 GeV.} 
\begin{tabular}{cccccc}
\\
     \hline     
     \hline
           &  & \multicolumn{3}{c}{} &   \\         
 elastic profile & $\sqrt{<b^2>_{tot}}$ &   
     \multicolumn{3}{c}{\hspace*{0.5cm}$\sqrt{<b^2>_{el}} \; \;$  } 
    & {$\sqrt{<b^2>_{inel}}$  }  \\  \\
    \cline{1-6}     
      &      & \hspace*{0.1cm} modulus & \hspace*{0.1cm} phase & 
      \hspace*{0.2cm} sum     &       \\  
      &[fm]  &   [fm]       & [fm]      & [fm]& [fm]      \\  
     \hline     
     \hline   \\
       peripheral& 1.033 & 0.676 &  1.671 
     &1.803 &   0.772    \\   \\
  \cline{1-6}  \\
    central    & 1.028 & 0.679 & $\sim$ 0. & 0.679 & 1.087   \\  \\ 
  \cline{1-6}
     \hline     
     \hline
\end{tabular}
\end{center}
\end{table}
It may be expected that the total profile 
entering into modified unitarity condition (\ref{pr2}) 
should be approximately of Gaussian type with the 
values that may be characterized by integral cross sections
and by RMS values shown in Table 1. The elastic profile will 
remain unchanged. Under these assumptions the total profile 
shape can be defined (the $s$ dependence being dropped) as
$h_{tot}(b) \; = \; a e^{- \beta b^2}$.
Using formulas 3.461 from Ref. \cite{grad} the
corresponding integrals needed for calculation
of the total cross section and total mean squared 
value can be analytically determined as 
\begin{equation}
\int \limits_{0}^{\infty} b \; db \;a \;e^{- \beta b^2} \; = \;
{a \over {2 \beta}},
\hspace*{1cm}
\int \limits_{0}^{\infty} b^3\; db\; a \;e^{- \beta b^2} \; = \;
{a \over {2 \beta^2}}
\label{pr4}
\end{equation}
and the values of the constants $a$ and $\beta$ 
can be determined from experimentally established values. 
For the peripheral case of elastic $pp$ scattering at energy 
of 53 GeV their values are: $a = 0.324$ and $\beta = 0.946$. 
\begin{table}
\begin{center}
\caption{The values of integrated cross sections and 
of the total, elastic and inelastic RMS. }
\begin{tabular}{cccc}  
\\ \hline
\hline  \\ 
 Quantity &\hspace*{2.8cm} &  Original values \hspace*{0.3cm} &  New values \\
 \\
\hline  \\
$\sigma_{tot}  $ &[mb]  &    42.864        &  42.872     \\
$\sigma_{el}\; $ &[mb]  &     7.479        &  7.479      \\
$\sqrt{<b^2>_{tot}}\;\;\;\;\;\; $ &[fm]$\;\;$  &     1.0   &   1.028   \\
$\sqrt{<b^2>}_{el} \;\;\;\;\;\; $ &[fm]$\;\;$  &     1.803 &   1.803   \\
$\sqrt{<b^2>}_{inel}\;\;\;\;\;\;$ &[fm]$\;\;$  &     0.772 &   0.772   \\  \\     
\hline \\
$\int \limits_{0}^{\infty} bdb\;c(s,b)$   & [fm$^2$]   & - & 0.029 \\
$\int \limits_{0}^{\infty} b^3db\;c(s,b)$ & [fm$^4$]   & - & 0.097 \\ \\
\hline
\hline
\end{tabular}
\end{center}
\end{table}
The $b$ dependence of the auxiliary function $c(s,b)$ is then 
determined with the help of the first equation from 
(\ref{pr1}) where $\Im h_1(s,b)$ is taken from experimental
analysis. And the second equation determines then the 
shape of the inelastic profile. 

Knowing the shapes of the total and inelastic profiles
together with the $b$ dependence of the auxiliary function 
$c(s,b)$ the values of all the integrated cross sections and
of all the mean squares can be verified numerically as can
be seen from the Table 2. The new values are practically 
quite comparable with the original ones. Also the values 
of the integrals of function $c(s,b)$ (see Eqs. (\ref{rs1}))
only slightly different from zero are shown in Table 2.
The modified profiles are exhibited in Fig. 4. The new total 
and inelastic profiles are central while the elastic profile
remains unchanged and is peripheral.

\section{Conclusion}
\label{sec4}
Some results concerning elastic nucleon collisions 
at high energies and based on the validity of optical 
theorem have been summarized; the eikonal model
has been applied to. The approach suitable for the 
case of finite energies has been presented. It has 
been shown that elastic processes may be interpreted 
as peripheral.

\end{document}